\documentclass[%
reprint,
superscriptaddress,
nofootinbib,  
amsmath,amssymb,
aps,
longbibliography
]{revtex4-2}

\usepackage{tabularx}
\usepackage{graphicx}
\usepackage{dcolumn}
\usepackage{bm}
\usepackage{xcolor}
\usepackage{upgreek}

\usepackage{appendix}

\newcommand{\im}{{i\mkern1mu}}    

\graphicspath{{Figures/}}

\begin{document}

\preprint{APS/123-QED}


\title{Performance enhancement in Josephson traveling wave parametric amplifiers by tailoring the relative distance between junctions}  

\author{M.A. Gal\'i Labarias}
\email{galilabarias.marc@aist.go.jp}
\affiliation{%
Global Research and Development Center for Business by Quantum-AI Technology (G-QuAT), National Institute of Advanced Industrial Science and Technology (AIST), Tsukuba, Ibaraki 305-8568, Japan
}
\author{T. Yamada}
\affiliation{%
Global Research and Development Center for Business by Quantum-AI Technology (G-QuAT), National Institute of Advanced Industrial Science and Technology (AIST), Tsukuba, Ibaraki 305-8568, Japan
}
\author{Y. Nakashima}%
\affiliation{%
Global Research and Development Center for Business by Quantum-AI Technology (G-QuAT), National Institute of Advanced Industrial Science and Technology (AIST), Tsukuba, Ibaraki 305-8568, Japan
}
\author{Y. Urade}%
\affiliation{%
Global Research and Development Center for Business by Quantum-AI Technology (G-QuAT), National Institute of Advanced Industrial Science and Technology (AIST), Tsukuba, Ibaraki 305-8568, Japan
}
\author{J. Claramunt}%
\affiliation{%
Universidad Carlos III de Madrid, Av. de la Universidad 30, 28911 Leganés, Madrid, Spain
}
\author{K. Inomata}
\email{kunihiro.inomata@aist.go.jp}
\affiliation{%
Global Research and Development Center for Business by Quantum-AI Technology (G-QuAT), National Institute of Advanced Industrial Science and Technology (AIST), Tsukuba, Ibaraki 305-8568, Japan
}




\date{\today}

\begin{abstract}
Josephson traveling wave parametric amplifiers with heterogeneously spaced junctions are theoretically investigated.
We consider unit cells with three junctions and characterize their interaction by spatially displaced fields defined by node fluxes. 
To solve this problem we define the system action and apply the variational principle to obtain the static action, which determines the equations of motion.
This work shows that gain and bandwidth can be increased by modifying the relative distance between junctions within the same unit cell, thus increasing the effective nonlinear interaction.
We find an optimal sub-unit-cell size ratio, which maximizes both gain and bandwidth, while equally spaced junctions offer minimum performance.
Even though this method does not rely on phase-matching, the same device can operate at different pump frequencies, it requires a very large number of JJs to achieve gains above 20 dB. We show that this method can be combined with resonant-phase matching and predicts an ideal gain above 29 dB on a 4 GHz bandwidth with 1998 junctions.
Despite possible challenges fabricating devices at the optimal sub-unit-cell size with current technologies, substantial gain increase can still be achieved at sub-optimal sizes.

\end{abstract}

\keywords{JTWPA, quantum computing, non-linear amplifiers, superconductors}
\maketitle

\section{\label{sec:Intro}Introduction}

%

Josephson Parametric Amplifiers~\cite{Yurke1989, Yamamoto2008, Castellanos2007, Bergeal2010, Mutus2014, Aumentado2020} have shown high potential as single-photon detectors~\cite{Mallet2011}, quantum-state generators~\cite{Yurke1988, Castellanos2008, Eichler2014} and highly-sensitive qubit readout systems~\cite{Lin2013, Hacohen2018, Planat2019}.
However, due to their resonant nature they are intrinsically narrow banded.
Josephson Traveling Wave Parametric Amplifiers (JTWPAs)~\cite{Cullen1958, Cullen1960, Sweeny1985} arise to overcome this handicap and nowadays represent a cornerstone component of superconducting quantum computers~\cite{Devoret2013, Esposito2021,Shiri2024}. These nonlinear quantum amplifiers are capable of reaching quantum-limited signal processing~\cite{HoEom2012, Yaakobi2013, Zorin2016, Macklin2015, Bell2015, Roudsari2023}, offering unrivaled potential for broadband, high-gain~\cite{Tien1958, Chen1989} and low-noise amplification~\cite{Caves1982, Houde2019}, and generation of quantum states~\cite{Grimsmo2017, Boutin2017, Perelshtein2022}.
Their fundamental element is the Josephson junction (JJ) which acts as the nonlinear element responsible for the signal amplification. Fabricating and designing these devices have many challenges, including: phase-matching~\cite{OBrien2014, Macklin2015, White2015, Planat2020, Peng2022_PRX, Rizvanov2024}, loss reduction~\cite{Esposito2021, Sandbo2025arx}, directionality~\cite{Qiu2023}, impedance matching~\cite{Ranadive2022}, pump-depletion~\cite{Zorin2019, Nikolaeva2023}, to name a few.
While many of these are currently being experimentally investigated there are still many fundamental questions that need to be explored and can give us a solution to current optimization challenges. 
For instance, theoretical understanding of the interplay between the nonlinear dynamics of Josephson junctions and wave propagation phenomena, optimizing phase-matching techniques for broadband operation and modeling the impact of quantum noise and higher-order nonlinear elements on amplifier performance~\cite{Nilsson2023, Dixon2020}.

This work presents a theoretical research on how the relative distance between Josephson junction affects the nonlinear interaction and energy exchange between modes.
To approach this problem we include more than one JJ per unit cell~\footnote{A unit cell is the smallest building block of the system.} and relate their interaction by spatially displaced fields, then we apply the variational principle to minimize the system action and obtain the JTWPA equations of motion. 
Using this method, we can recover the usual differential equations of a typical JTWPA when all the JJs in the unit cell are identical and their relative distance equal, thus we can smoothly transition from a typical unit cell to a more dense one.

The outcomes of this theoretical investigation will not only inform experimental developments, but also contribute to a broader understanding of nonlinear wave dynamics in superconducting circuits, a topic with profound implications for both classical and quantum technologies.

\section{\label{sec:Math}Mathematical model}
\begin{figure}
	\centering
	\includegraphics[width=\linewidth]{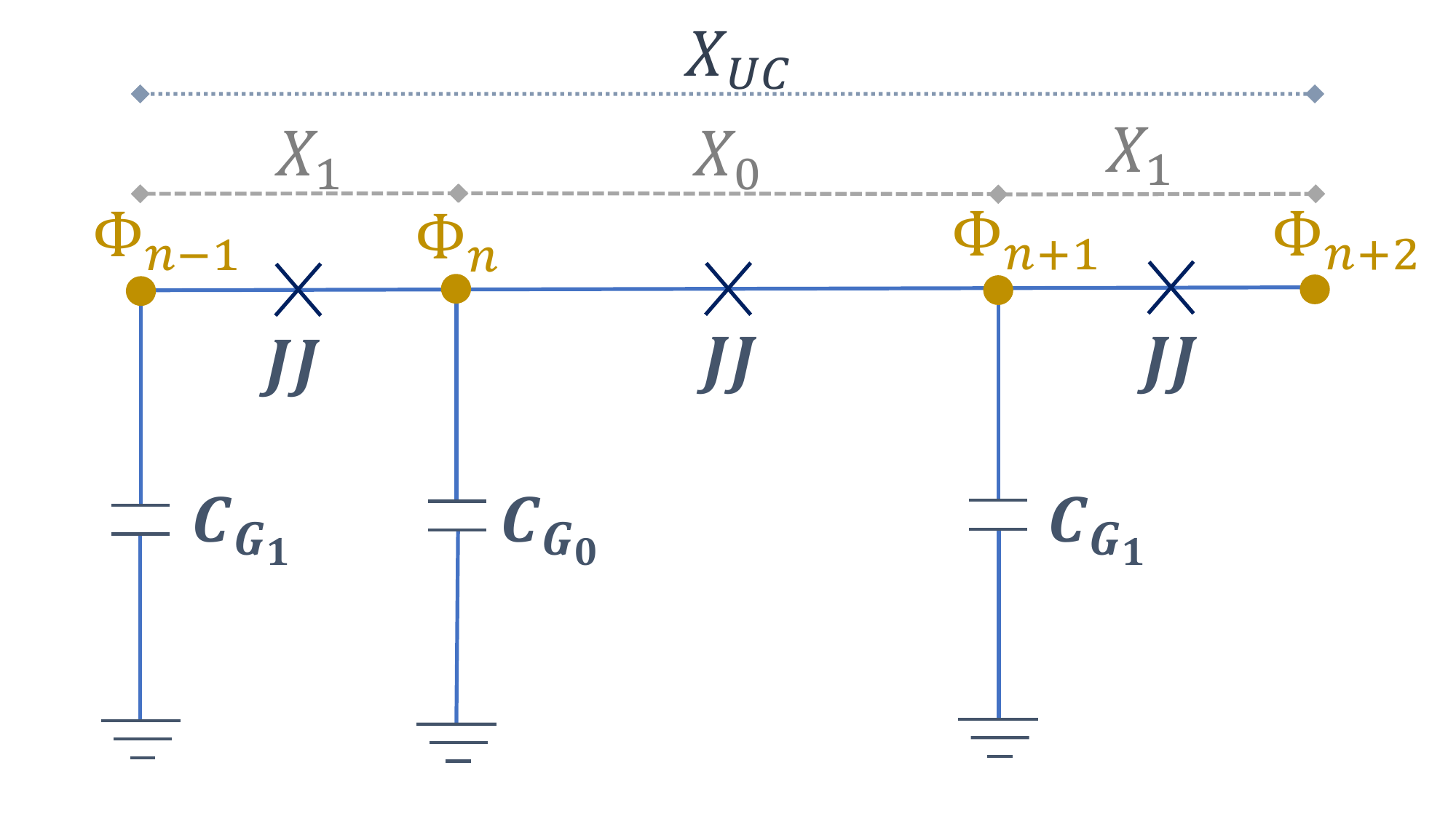}
	\caption{Circuit diagram of the \emph{Sando} unit cell with three Josephson junctions. The total unit cell length is $X_{UC}$ and each sub-unit cell has a length defined by $X_{0,1}$. The crosses represent Josephson junctions, composed by a nonlinear inductance and capacitance, and each sub-unit cell contains a ground capacitance $C_{G_{0,1}}$. The gold circles indicate the node fluxes $\Phi_n$~\cite{Devoret1995} between sub-unit cells.}
	\label{fig:unit_cell}
\end{figure}

In superconducting electronics, usually the scale of the circuit's unit cell is significantly smaller than the wavelength of interest, therefore continuous approximations are usually applied to model these systems~\cite{Yaakobi2013,Yaakobi2013Err, OBrien2014} and have been experimentally  validated~\cite{Macklin2015, Sandbo2025arx}. Here, we study the case of having several JJs within the unit cell and how their relative distances affect JTWPA performance. To do so, we investigate the simple case of a three junction unit cell, where the first and last sub-unit-cell sizes are equal (see Fig.~\ref{fig:unit_cell}). We will refer to this kind of unit cell as \emph{Sando} unit cell (from the Japanese pronunciation of sandwich).

We start by defining the Lagrangian density using the unit cell kinetic and potential terms, i.e. ${\mathcal{L} = \mathcal{T} - \mathcal{U}}$. For the unit cell described in Fig.~\ref{fig:unit_cell}, the discrete Lagrangian density can be expressed as follows,
\begin{widetext}
\begin{align}\label{eq:sando_Lagrangian}
\tilde{\mathcal{L}}(\phi, \partial_{\tau} \phi) &= E_{J} \Big\{ 
\frac{c_{G_0}}{2} \left( \partial_{\tau}  \phi_n(\tau) \right)^2 
+ \frac{c_{G_1}}{2}   \left( \partial_{\tau}  \phi_{n-1}(\tau) \right)^2
+ \frac{c_{G_1}}{2}  \left( \partial_{\tau}  \phi_{n+1} (\tau) \right)^2 
+ \frac{1}{2} \Big(  \left(\Delta  \partial_{\tau}  \phi_{n-1}(\tau)   \right)^2 \nonumber \\
&+  \left( \Delta  \partial_{\tau}  \phi_n(\tau) \right)^2 
+ \left( \Delta \partial_{\tau}  \phi_{n+1}(\tau) \right)^2  \Big)
+  \cos \left( \Delta \phi_{n-1}(\tau) \right) + \cos\left( \Delta \phi_n (\tau)\right) + \cos\left( \Delta \phi_{n+1} (\tau)\right) \Big\} \, ,
\end{align}
\end{widetext}
where $\Delta \phi_n := \phi_{n+1}  - \phi_n$ and small case letters indicate normalized variables, such that: ${c_{G_{0,1}} := C_{G_{0,1}}/C_J}$ with $C_J$ the JJ capacitance, 
and ${ \phi_n := \Phi_n/\Phi_0}$ with $\Phi_0$ the flux quantum.
The Josephson energy is $E_J = I_c \Phi_0/(2\pi) $ with $I_c$ the JJ critical current (here we are assuming identical JJs across the device). We have also introduced the normalized time $\tau := \Omega_J t$ where $\Omega_J =\left( L_J C_J \right)^{-1/2}$ is the Josephson plasma frequency and ${L_J = \Phi_0 / (2\pi I_c)}$ the Josephson inductance.

Next, assuming that the unit-cell size is smaller than the wavelengths under study, we are entitled to apply the continuous approximation to the spatial coordinates:
\begin{subequations}\label{eq:cont_appr}
\begin{align}
\Delta \phi_{n-1} (\tau)& \approx x_1  \partial_x \phi \left( x - a , \tau \right) \, , \\
\Delta \phi_{n} (\tau)& \approx x_0  \partial_x \phi(x, \tau) \, , \\
\Delta \phi_{n+1} (\tau)&\approx x_1  \partial_x \phi \left( x + a , \tau\right) \, , 
\end{align}
\end{subequations}
where we have used central finite differences and normalized the spatial variable by the unit-cell length ${X_{UC} = X_0+2X_1}$, i.e. $x_{0,1} := X_{0,1}/X_{UC}$. To simplify notation we have introduced $a := (x_0 + x_1)/2$.

Using Eqs.~\eqref{eq:sando_Lagrangian} and \eqref{eq:cont_appr}, we obtain the continuous Lagrangian density:
\begin{widetext}
\begin{align}\label{eq:sando_Lagrangian2}
\mathcal{L}&(\phi, \partial_{\tau} \phi) = E_{J} \Big\{ 
\frac{c_{G_0}}{2} \left( \partial_{\tau}  \phi(x, \tau) \right)^2 
+ \frac{c_{G_1}}{2}  \left( \partial_{\tau}  \phi(x -a, \tau) \right)^2
+ \frac{c_{G_1}}{2}  \left( \partial_{\tau}  \phi(x + a, \tau) \right)^2  
+ \frac{1}{2} \left( x_1\partial_x \partial_{\tau} \phi(x - a, \tau) \right)^2 \nonumber \\ 
&+ \frac{1}{2}  \left( x_0 \partial_x \partial_{\tau} \phi(x, \tau)  \right)^2  
+ \frac{1}{2} \left( x_1\partial_x \partial_{\tau} \phi(x + a, \tau)  \right)^2 
+  \cos \left( x_1 \partial_x \phi (x - a, \tau)  \right) + \cos \left( x_0\partial_x \phi (x, \tau) \right)
+ \cos \left( x_1\partial_x \phi (x + a, \tau)  \right) \Big\} \, .
\end{align}
\end{widetext}

The Lagrangian density shown in Eq.~\eqref{eq:sando_Lagrangian2} contains spatially displaced fields, i.e. $\phi(x \pm a, \tau)$, thus applying the Euler-Lagrange equations is not straight forward.
Instead, to obtain the system's equations of motion, here we use the variational principle directly and find which trajectory minimizes the action of our Lagrangian.
The action of the system is defined from the Lagrangian density as,
\begin{equation}\label{eq:S_def}
    S = \iint dxd\tau \, \mathcal{L} \, .
\end{equation}

Minimizing the action means imposing that the variation of the action is zero, i.e. $\delta S=0$, so the integral equation that we must solve is
\begin{equation}\label{eq:iintL}
    \iint dxd\tau \, \delta \mathcal{L}  = 0 \, .
\end{equation}

Before applying the variational principle, we need to define properly the field $\phi(x, \tau)$.
To do so, we note that $\phi(x, \tau)$ can be expressed using the Fourier-transform operators as,
\begin{align}
	\phi(x, \tau) = \frac{1}{\sqrt{2\pi}}\int_{-\infty}^{+\infty} d\omega \,  \hat{\phi}(x, \omega) e^{-\im \omega \tau} \, ,
    \label{eq:mode_Fourier}
\end{align}
where for later convenience and without loss of generality, we express the Fourier transform as follows,
\begin{align}
	\hat{\phi}(x, \omega) &= \frac{1}{2} A(x, \omega) e^{\im k (\omega) x } \, ,	 \label{eq:An_Fourier}
\end{align}
with $k(\omega)$ the wavenumber.
Then the relationship between the time-dependent field, $\phi(x, \tau)$, and $\hat{\phi}(x, \omega)$ ( or $A(x, \omega)$) can be obtained by applying the inverse Fourier transform to Eq.~\eqref{eq:mode_Fourier}.
Now, we can express the Fourier transform of the $\pm a$ displaced field in terms of the non-displaced field using Eq.~\eqref{eq:An_Fourier},
\begin{align}
	\hat{\phi}(x\pm a, \omega) &\approx \hat{\phi}(x, \omega)  e^{ \pm \im k(\omega) a } \, ,	 \label{eq:phi_an}
\end{align}
where we are taking the \emph{Slowly Varying Envelope} (SVE) approximation, i.e. ${|A(x, \omega)| \approx |A(x \pm a, \omega)|}$.
We have introduced all the relationships that we need to apply the variational principle to our Lagrangian density, Eq.~\eqref{eq:sando_Lagrangian2}, which contains displaced fields.

To make the notation more treatable, we will express the Lagrangian density in three terms,
\begin{equation}
    \mathcal{L}  = \mathcal{L}_1 + \mathcal{L}_2 + \mathcal{L}_3 \, .
\end{equation}

Let us start with
\begin{align}
    \mathcal{L}_1 =& \frac{E_J}{2} \Big[  c_{G_0} \dot{ \phi}(x, \tau)^2 
    + c_{G_1}  \dot{ \phi}(x - a, \tau)^2 \nonumber \\&
    + c_{G_1}  \dot{ \phi}(x + a, \tau)^2 \Big]\, ,
\end{align}
where for readability purposes, we are using the notation ${\dot{\phi} := \partial_{\tau} \phi}$. Then,

\begin{align}
    \delta \mathcal{L}_1 &= E_J \Big[ c_{G_0} \dot{ \phi}(x, \tau) \, \delta \dot{ \phi}(x, \tau) \nonumber \\
    &+ c_{G_1}  \dot{ \phi}(x - a, \tau) \, \delta \dot{ \phi}(x - a, \tau) \nonumber \\
    &+ c_{G_1}  \dot{ \phi}(x + a, \tau) \, \delta \dot{ \phi}(x + a, \tau) \Big] \, .
\end{align}

To compute the variation of the displaced fields as variations of the fields without displacement, we adhere to the following procedure:
\begin{enumerate}
	\item Express the fields $\phi(x+\epsilon a,\tau)$, $\epsilon \in \{-1,0,1\}$, in their Fourier expansion using Eq.~\eqref{eq:mode_Fourier}.
	\item Then, we apply the approximation Eq.~\eqref{eq:phi_an}.
	\item Finally, we apply the variation to express ${\delta \phi(x + \epsilon a,\tau)}$ in terms of $\delta \hat{\phi}(x,\omega)$.
\end{enumerate}


Using these steps,

\begin{align}
    \delta \mathcal{L}_1  =& -\frac{E_J}{2\pi}  \iint d\omega d\omega'\,   \omega \, \omega' \clubsuit (\omega, \omega') \hat{\phi}(x, \omega) \nonumber \\ &
    \times e^{-\im (\omega' + \omega )\tau} \, \delta \hat{\phi}(x, \omega')  \, ,
\end{align}
where 
\begin{align}
\clubsuit (\omega , \omega') :=   c_{G_0} + 2 c_{G_1} \cos \big[ k(\omega) a + k(\omega') a \big] \, . 
\end{align}

We can now obtain the variation of the action corresponding to $\mathcal{L}_1$ by applying Eq.~\eqref{eq:S_def},
\begin{align}\label{eq:dS_1}
    \delta S_1 = & \iint dxd\tau \, \delta \mathcal{L}_1 \nonumber \\
    = & - E_J \iiint dx d\omega d\omega'\,   \omega \, \omega' \clubsuit (\omega, \omega') \hat{\phi}(x, \omega) \nonumber \\ &
    \times \delta \hat{\phi}(x, \omega') \frac{1}{2\pi} \int d\tau e^{-\im (\omega' + \omega )\tau} \\
     = & - E_J\iiint dx d\omega d\omega'\,   \omega \, \omega' \clubsuit (\omega, \omega')\hat{\phi}(x, \omega) \nonumber \\ &
    \times \delta \hat{\phi}(x, \omega') \, \delta_{\text{D}} \left(\omega' + \omega \right) \\
    = & E_J\iint dx d\omega \,   \omega^2 \clubsuit (\omega, -\omega)\phi(x, \omega) \nonumber \\ &
    \times \delta \hat{\phi}(x, -\omega) \\
    = & \frac{E_J}{\sqrt{2\pi}} \iiint dx d\tau d\omega \,   \omega^2 \clubsuit (\omega, -\omega)\hat{\phi}(x, \omega) \nonumber \\ &
    \times e^{-\im \omega \tau} \, \delta \phi(x, \tau)
    \, ,
\end{align}
where we have used the properties of the \emph{delta distribution}, $\delta_{\text{D}} \left(\omega' + \omega \right)$, which couples the $\omega$ and $-\omega'$ modes, and in the last step we applied the inverse Fourier transform to recover the time dependence of the variational field.

The second term on the Lagrangian density is,
\begin{align}
    \mathcal{L}_2 =&  \frac{E_J}{2} \Big\{ x_0^2\left[ \partial_x\dot{\phi}(x, \tau)\right]^2 + x_1^2\left[ \partial_x\dot{\phi}(x - a, \tau)\right]^2 \nonumber \\ &
    + x_1^2\left[ \partial_x\dot{\phi}(x + a, \tau)\right]^2 \Big\}
     \, ,
\end{align}
then its variation is,
\begin{align}\label{eq:delta_L2}
    \delta \mathcal{L}_2 =& E_J \Big[ x_0^2 \partial_x\dot{\phi}(x, \tau) \, \delta  \partial_x\dot{\phi}(x, \tau) \nonumber \\ &
    + x_1^2 \partial_x\dot{\phi}(x - a, \tau) \, \delta \partial_x\dot{\phi}(x - a, \tau) \nonumber \\ &
    + x_1^2 \partial_x\dot{\phi}(x + a, \tau) \, \delta \partial_x\dot{\phi}(x + a, \tau)  \Big]
     \, .
\end{align}

To simplify Eq.~\eqref{eq:delta_L2}, we first notice that the variation and partial derivative commute by definition, so we can rewrite Eq.~\eqref{eq:delta_L2} as
\begin{align}\label{eq:delta_L2b}
    \delta \mathcal{L}_2 =& - E_J \Big[ x_0^2 \partial_x^2\dot{\phi}(x, \tau) \, \delta \dot{\phi}(x, \tau) \nonumber \\ &
    + x_1^2 \partial_x^2\dot{\phi}(x - a, \tau) \, \delta \dot{\phi}(x - a, \tau) \nonumber \\ &
    + x_1^2 \partial_x^2\dot{\phi}(x + a, \tau) \, \delta \dot{\phi}(x + a, \tau) 
    \Big] 
    + \frac{\partial  \mathcal{R}_2}{\partial x}
    \, .
\end{align}
where Eq.~\eqref{eq:delta_L2} and Eq.~\eqref{eq:delta_L2b} are equal up to a total derivative $ \frac{\partial \mathcal{R}_2}{\partial x}$.
When calculating the action by integrating Eq.~\eqref{eq:delta_L2b}, the total derivative term can be evaluated easily, and only depends on the boundary conditions. Here, we are assuming negligible boundary conditions~\footnote{For this study we will only consider the simpler case of forward propagating modes, if ones is interested in interface reflections and the effects of backward propagating modes, then adding the boundary conditions is necessary. This does not change the theory presented here, but adds constrains to the initial conditions of the system of differential equations defining the JTWPA.}, so without loss of generality we can omit this term in the following derivation.

Using the same method as before, Eq.~\eqref{eq:delta_L2b} can be expressed as
\begin{align}
    \delta \mathcal{L}_2 =&  \frac{E_J}{2\pi} \iint d\omega d\omega'  \omega \, \omega' \heartsuit (\omega, \omega') \partial_x^2 \hat{\phi}(x, \omega)  \nonumber \\ &
    \times e^{-\im (\omega + \omega')\tau} \, \delta  \hat{\phi} (x, \omega')   
    \, .
\end{align}
where
\begin{align}
\heartsuit (\omega, \omega') := & x_0^2   + 2 x_1^2 \cos \big[ k(\omega)a + k(\omega') a \big] \, .
\end{align}

Following the same method, we can obtain the variation of the action corresponding to $\mathcal{L}_2$,
\begin{align}\label{eq:dS_2}
    \delta S_2 =&  -\frac{E_J}{\sqrt{2\pi}} \iiint dx d\tau d\omega \, \omega^2 \heartsuit (\omega, -\omega) \partial_x^2 \hat{\phi}(x, \omega)  \nonumber \\ &
    \times e^{-\im \omega \tau} \, \delta \phi (x, \tau)
    \, .
\end{align}

Using similar methods, the variation of last term of the Lagrangian density,
\begin{align}\label{eq:L_3}
	\mathcal{L}_3  =& E_{J} \Big[ \cos \left( x_0\partial_x \phi (x, \tau) \right) +
	 \cos \left( x_1 \partial_x \phi (x - a, \tau)  \right)
	\nonumber \\&
	+ \cos \left( x_1\partial_x \phi (x + a, \tau)  \right) \Big] \, ,
\end{align}
can be expressed as,
\begin{align}\label{eq:deltaL_3a}
	&\delta \mathcal{L}_3  = -E_J\Big[
	\sin \left[ x_0 \partial_x  \phi (x , \tau ) \right]x_0  \, \delta  \partial_x \phi (x , \tau ) \nonumber \\    &
	+ \sin \left[ x_1 \partial_x  \phi (x - a, \tau ) \right]x_1  \, \delta  \partial_x \phi (x - a, \tau ) \nonumber \\ &
	+ \sin \left[ x_1 \partial_x  \phi (x + a, \tau ) \right]x_1  \, \delta  \partial_x \phi (x + a, \tau ) \Big]
	\, .
\end{align}

Similar than before, $\delta \mathcal{L}_3$ can be expressed as
\begin{align}\label{eq:deltaL_3b}
	&\delta \mathcal{L}_3  = E_J\Big\{
	\cos \left[ x_0 \partial_x  \phi (x , \tau ) \right]x_0^2 \partial_x^2  \phi (x , \tau ) \, \delta \phi (x , \tau ) \nonumber \\    &
	+ \cos \left[ x_1 \partial_x  \phi (x - a, \tau ) \right]x_1^2 \partial_x^2  \phi (x - a, \tau ) \, \delta \phi (x - a, \tau ) \nonumber \\ &
	+ \cos \left[ x_1 \partial_x  \phi (x + a, \tau ) \right]x_1^2 \partial_x^2  \phi (x + a, \tau ) \, \delta  \phi (x + a, \tau ) \Big\}
	\nonumber \\ &
	+ \frac{\partial \mathcal{R}_3}{\partial x}
	\, ,
\end{align}
where Eq.~\eqref{eq:deltaL_3a} and Eq.~\eqref{eq:deltaL_3b} are equal up to a total derivative $ \frac{\partial \mathcal{R}_3}{\partial x}$. Similarly to Eq.~\eqref{eq:delta_L2b}, the total derivative term will be omitted for simplicity.

Applying Taylor expansion up to second order we obtain the usual nonlinear terms corresponding to four-wave mixing,
\begin{align}
	\delta  \mathcal{L}_3  =& \frac{E_J}{2\pi} \int d\omega' \Big\{ \int d\omega \, \diamondsuit^{\text{L}} (\omega, \omega') \partial_x^2 \big[ \hat{\phi}(x, \omega) \big] e^{-\im (\omega + \omega')\tau} \nonumber \\
	&- \frac{1}{2} \iiint d\omega_1 d\omega_2 d\omega_3\diamondsuit^{\text{NL}}(\omega_1, \omega_2, \omega_3, \omega') \nonumber \\ &
	\partial_x^2 \big[ \hat{\phi}(x, \omega_1) \big] \partial_x \big[ \hat{\phi}(x, \omega_2) \big] \partial_x \big[ \hat{\phi}(x, \omega_3) \big] \nonumber \\
	& \times e^{-\im \left( \omega_1 + \omega_2 + \omega_3 + \omega'\right) \tau}\Big\} \, \delta  \hat{\phi}(x, \omega')  \, ,
\end{align}
where we have defined
\begin{align}
	& \diamondsuit^{\text{L}}(\omega, \omega') :=  x_0^2 + 2 x_1^2 \cos \big[ k(\omega)a + k(\omega')a \big]  \, , \\
	& \diamondsuit^{\text{NL}}(\omega_1, \omega_2, \omega_3, \omega_4) := x_0^4 + 2 x_1^4 \cos \left[ \sum_{n=1}^4 k(\omega_n) a \right] \, .
\end{align}

Then, the variation of the action corresponding to $\mathcal{L}_3$ is
\begin{align}\label{eq:dS_3}
	\delta  S_3  &= \frac{E_J}{\sqrt{2\pi}}\iint dx d\tau \Big\{ \int d\omega \, \diamondsuit^{\text{L}} (\omega, -\omega) \partial_x^2 \big[ \hat{\phi}(x, \omega) \big] e^{-\im \omega \tau} \nonumber \\
	&- \frac{1}{2} \iiint d\omega_1 d\omega_2 d\omega_3\diamondsuit^{\text{NL}}\big(\omega_1, \omega_2, \omega_3, -\sum_{n=1}^3 \omega_n \big) \nonumber \\ &
	\partial_x^2 \big[ \hat{\phi}(x, \omega_1) \big] \partial_x \big[ \hat{\phi}(x, \omega_2) \big] \partial_x \big[ \hat{\phi}(x, \omega_3) \big] \nonumber \\
	& \times e^{-\im \left( \omega_1 + \omega_2 + \omega_3 \right) \tau}\Big\} \, \delta  \phi(x, \tau)  \, .
\end{align}

We can now find the static action by solving Eq.~\eqref{eq:iintL} using Eqs.~\eqref{eq:dS_1}, \eqref{eq:dS_2} and \eqref{eq:dS_3}, which leads us to the following integral equation,
\begin{widetext}
\begin{align}\label{eq:action_DE}
 & \int d\omega \,\clubsuit (\omega, -\omega) \omega^2 \hat{\phi}(x, \omega) e^{-\im \omega \tau}
  - \Big[ \omega^2  \heartsuit(\omega, -\omega) - \diamondsuit^{\text{L}} (\omega, -\omega) \Big] \Big[ 2\im k(\omega) \partial_x \big[\hat{\phi}(x, \omega) \big]  - k(\omega)^2 \hat{\phi}(x, \omega)\Big]   e^{-\im \omega \tau}  \nonumber \\ 
 &=\frac{1}{2} \iiint d\omega_1 d\omega_2 d\omega_3 \, \diamondsuit^{\text{NL}}\big(\omega_1, \omega_2, \omega_3, -\sum_{n=1}^3 \omega_n \big) \partial_x^2 \big[ \hat{\phi}(x, \omega_1) \big] \partial_x \big[ \hat{\phi}(x, \omega_2) \big] \partial_x \big[ \hat{\phi}(x, \omega_3) \big] e^{-\im \left( \omega_1 + \omega_2 + \omega_3 \right) \tau}  \, ,
\end{align}

\end{widetext}
where we have assumed that $|\partial_x^2 \hat{\phi}| \ll |\partial_x \hat{\phi}| \ll |\hat{\phi}|$, and approximated
\begin{equation}
\partial_x^2 \hat{\phi}(x, \omega) \approx  2\im k(\omega) \partial_x \hat{\phi}(x, \omega)   - k(\omega)^2 \hat{\phi}(x, \omega)    \, .
\end{equation}

Equation~\eqref{eq:action_DE} can be simplified by defining the wavenumber $k(\omega)$ to be
\begin{align}
 \omega^2 \clubsuit (\omega)   + k(\omega)^2 \Big[ \omega^2  \heartsuit(\omega) - \diamondsuit^{\text{L}} (\omega)\Big] = 0 \, ,
\end{align}
i.e. equivalent to a linear dispersion relation.

Then,
\begin{equation}\label{eq:k_sando}
 k(\omega) =  \frac{ \omega  }{ \sqrt{ 1 -  \frac{\heartsuit(\omega)}{\diamondsuit^{\text{L}} (\omega)}\omega^2 } }\sqrt{\frac{\clubsuit  (\omega) }{\diamondsuit^{\text{L}}  (\omega) }}
  \, .
\end{equation}

In the present work, we are only interested in forward propagating modes; thus, in Eq.~\eqref{eq:k_sando}, we have chosen the positive sign of the square-root, which gives the same parity between $\omega$ and $k$, i.e. $k(\pm \omega) = \pm k(\omega)$. This choice also removes the frequency dependence of the linear scaling factors, which become
\begin{subequations}\label{eq:sando_L_factors}
\begin{align}
 \clubsuit = & c_{G_0} + 2 c_{G_1} \, , \\
 \heartsuit = & x_0^2 + 2 x_1^2  \, , \\
 \diamondsuit^{\text{L}} = & x_0^2 + 2 x_1^2  \, .
\end{align}
\end{subequations}

However, the nonlinear scaling factor still retains the modes dependence
\begin{align}\label{eq:sando_NL_factor}
	\diamondsuit^{\text{NL}}_{n;m_{1,2,3}} =& x_0^4 + 2 x_1^4 \cos \Big[ (k_{m_1} + k_{m_2} + k_{m_3} - k_n )a \Big]  \, .
\end{align}

We can see that for a \emph{Sando} unit cell with identical JJs the equality $\heartsuit=\diamondsuit^{\text{L}}$ holds.
Also, note that when sub-unit cells are equal, i.e. $x_0=x_1=1/3$ and $c_{G_0}=c_{G_1}$, we recover the usual wavenumber definition for a typical JTWPA unit cell.
However, $\diamondsuit^{\text{NL}}_{n;m_{1,2,3}}$ retains mode dependence, even if the sub-unit cells are identical (note that since $\cos \left[ (k_{m_1} + k_{m_2} + k_{m_3} - k_n )a \right] \approx 1$ this mode dependence is negligible as shown in Fig.~\ref{fig:gain_pumps}).

Using these definitions, Eq.~\eqref{eq:action_DE} can be discretized to a finite number of modes, $M$, as
\begin{align}\label{eq:sando_DE1}
  \frac{\partial A_n}{\partial x}  =&  - \frac{\im k_{n}}{2^4\omega_n^2 \clubsuit }  \sum_{m_{1,2,3}=-M}^M  \diamondsuit_{n;m_{1,2,3}}^{\text{NL}} \,  k_{m_1}^2 k_{m_2} k_{m_3} \nonumber \\
  &\times A_{m_1}  A_{m_2} A_{m_3} \; e^{ \im (k_{m_1}+ k_{m_2} + k_{m_3} - k_n ) x }   \, , 
\end{align}
with negative indexes corresponding to complex-conjugate modes, such that ${A_{-m}:=A_m^*}$, and the mode frequencies satisfying
\begin{align}
	\omega_n = \omega_{m_1}+ \omega_{m_2} + \omega_{m_3} \, , \label{eq:sando_DE2}
\end{align}
with $\omega_{-m} \equiv -\omega_m$ and the complex amplitudes $A_n$ defined by Eq.~\eqref{eq:An_Fourier}.

From Eq.~\eqref{eq:sando_DE2} we can see that this is a four-wave nonlinear process. Here, we consider three modes: pump, signal and idler; and use the four-wave mixing phase-matching condition for a degenerate pump $\omega_s + \omega_i = 2\omega_{p}$. At this point we can use typical approaches~\cite{Chen1989, Yaakobi2013, OBrien2014} to solve the system of coupled differential equations defined in Eq.~\eqref{eq:sando_DE1}.
Applying the SVE approximation, after some linear algebra we obtain the following equations,
\begin{subequations}\label{eq:sando_DE_final}
\begin{align}
 \frac{\partial \xi_{p}}{\partial x} &=  \frac{\diamondsuit_{\Delta k}^{\text{NL}}}{2^3\omega_{p}^2 \clubsuit} 
 k_{p}^2 k_s k_i \left(k_s + k_i - k_{p} \right) \xi_s \xi_i  \xi_{p} \sin (\Theta) \, , \label{eq:sando_DE_final_p} \\ 
 \frac{\partial \xi_s}{\partial x} &= \frac{\diamondsuit_{\Delta k}^{\text{NL}}}{2^4\omega_s^2\clubsuit} 
 k_{p}^2 k_s k_i \left(k_i - 2k_{p} \right) \xi_{p}^2 \xi_i \sin (\Theta) \, , \label{eq:sando_DE_final_s} \\
 \frac{\partial \xi_i}{\partial x} &= \frac{\diamondsuit_{\Delta k}^{\text{NL}}}{2^4\omega_i^2 \clubsuit} 
 k_{p}^2 k_s k_i \left(k_s - 2k_{p} \right) \xi_{p}^2 \xi_s \sin (\Theta) \, , \label{eq:sando_DE_final_i} \\
\frac{\partial \Theta}{\partial x} &= \Delta k_L + 2 \frac{\partial \theta_{p} }{\partial x} 
- \frac{\partial \theta_s }{\partial x} - \frac{\partial \theta_i }{\partial x} \, , \label{eq:sando_DE_final_th}
\end{align}
\end{subequations}
where we have expanded the normalize flux modes in real-valued functions as $A_n(x) = \xi_n(x) \exp(\im \theta_n(x))$
and defined ${\Theta  = \Delta k_L  x + 2\theta_{p}  - \theta_{s} -  \theta_{i}}$ with ${\Delta k_L = 2k_{p} - k_s - k_i }$ and $\diamondsuit_{\Delta k}^{\text{NL}}:=\diamondsuit_{n; m_{1,2,3}}^{\text{NL}}$ such that $k_{m_1} + k_{m_2} + k_{m_3} - k_n = \pm \Delta k_{L}$.
Equation~\eqref{eq:dthetadx} shows the expanded expression for $\partial \Theta /\partial x$, which also have dependence on the nonlinear scaling factor $\diamondsuit_{n; m_{1,2,3}}^{\text{NL}}$.
For identical sub-unit cells, or alternatively imposing $x_1=0$ and $c_{G_1}=0$, Eqs.~\eqref{eq:sando_DE_final} agree with the equations of a single junction unit cell model~\cite{Chen1989, OBrien2014}.

Eqs.~\eqref{eq:sando_DE_final} show that the nonlinear scaling factor $\diamondsuit_{n; m_{1,2,3}}^{\text{NL}}$, which has dependence on the modes frequencies, scales the energy exchange Eqs.~\eqref{eq:sando_DE_final_p} - \eqref{eq:sando_DE_final_i} and mode phases Eqs.~\eqref{eq:sando_DE_final_th} and \eqref{eq:dthetadx}. Since this nonlinear factor has modes dependence it cannot be absorbed as a constant proportional factor across the system of coupled differential equations. If that would be the case, it would prove that the relative spacing between JJs is irrelevant; however, as it is not the case, we can confirm that the relative spacing between JJs affects the nonlinear wave-mixing process.

The coupled differential equations can be solved numerically to analyze the different modes spatial evolution. However, for design and optimization purposes analytical expressions are extremely useful.
Applying the usual strong-pump approximation to Eqs.~\eqref{eq:sando_DE1} and the usual methods (see Appendix~\ref{app:strongpump}), we obtain the gain of the signal mode in decibels (dB):
\begin{equation}\label{eq:Sando_Gs}
G(x) = 10\log_{10}\left[ \cosh^2(gx) + \left(\frac{\Delta k}{2g} \right)^2 \sinh^2(gx) \right] \, .
\end{equation}

Equation~\eqref{eq:Sando_Gs} is the typical expression for the gain of a nonlinear amplifier~\cite{OBrien2014}, however in this case, the phase-difference, $\Delta k$ [Eq.~\eqref{eq:sando_Dk}], and gain coefficient, $g$ [Eq.~\eqref{eq:sando_g}], contain the scaling factors as seen in Eqs.~\eqref{eq:sando_gain_factors_a_p}, \eqref{eq:sando_gain_factors_a_si} and \eqref{eq:sando_gain_factors_b}.


\section{\label{sec:Dis}Discussion}

In this section, we investigate the \emph{Sando} unit cell for different sub-unit-cell sizes, total number of JJs, $N_{JJ}$, and signal and pump power.
In this paper, unless stated otherwise, we use the simulation input parameters defined in Tab.~\ref{tab:param}.
The initial conditions used to solve Eqs.~\eqref{eq:sando_DE_final} and Eq.~\eqref{eq:Sando_Gs} are ${I_p/I_c= 0.5}$ ($2.5 \, \upmu \text{A}$), $I_s/I_p =10^{-5}$ ($25 \,  \text{pA}$) and $I_i /I_p= 10^{-8}$ ($25 \, \text{fA}$).

The geometric components of the \emph{Sando} unit cell, Eqs.~\eqref{eq:sando_L_factors} and \eqref{eq:sando_NL_factor}, can be characterized by a single parameter, $x_0$, since by definition $x_1 = (1 - x_0)/2$ and $a = (1 + x_0)/4$. Thus, we refer to $x_0$ as the sub-unit-cell size and use this parameter to investigate the different \emph{Sando} sub-unit-cell size ratios.
\begin{table}[htb]
\begin{ruledtabular}
\begin{tabular}{c c c c c c}
    &&&&&\\[-.8em]
    $\; N_{JJ} \;$ & $\; I_p/I_c \; $ & $f_p$  & $I_c$ & $C_J$  & $C_{G_{0,1}} $  \\ [0.2ex] 
     & & (GHz) & ($\upmu$A)&  (fF)  &  (fF)  \\ [0.2ex] 
   \hline 
    &&&&&\\[-.8em]
   $1998$ &$0.5$ & $10 \;$  & $5 \; $ & $200\;$ & $26.3\;$ \\
\end{tabular} 
\end{ruledtabular}
\caption{Simulation parameters: $N_{JJ}$ is the number of Josephson junctions,  $I_p/I_c$ is the normalized pump current, $f_p$ is the pump frequency, $I_c$ is the JJ critical current and $C_{G_{0,1}}$ the sub-unit cell ground capacitances depicted in Fig.~\ref{fig:unit_cell}.}
\label{tab:param}
\end{table}

\begin{figure*}
	\centering
	\includegraphics{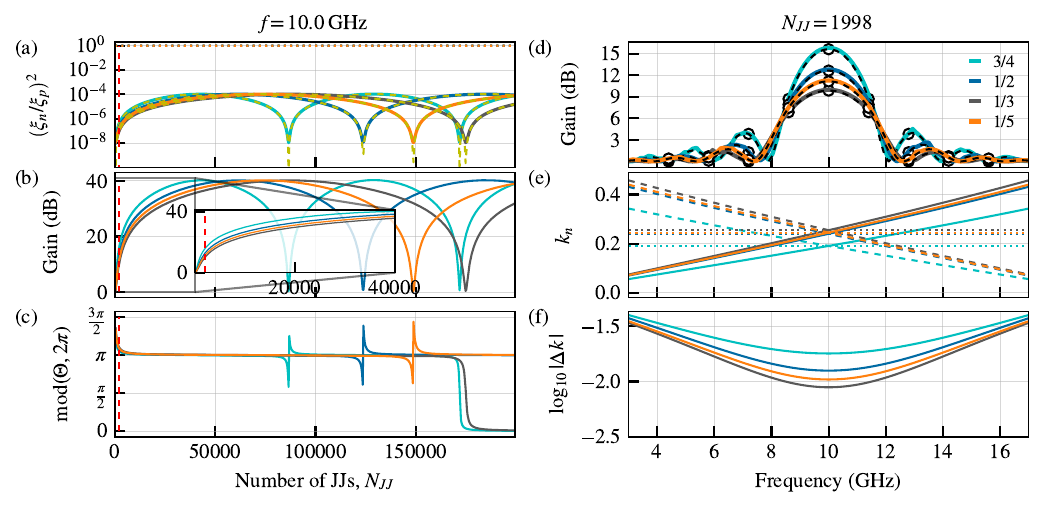}
	\caption{
		Study of the sub-unit-cell size's effect on the propagating modes of the \emph{Sando} JTWPA with four different sub-unit-cell sizes: $x_0 = 3/4$ (cyan lines), $x_0 = 1/2$ (blue lines), $x_0$ = 1/3 (gray lines) and $x_0 = 1/5$ (orange lines).
		(a) and (b) show the normalized powers of the signal (colored solid lines), idler (dashed yellow lines) and pumps (dotted lines) depending on the $N_{JJ}$ in  the JTWPAs: (a) logarithmic scale and (b) linear scale.
		(c) Total spatial dependent phase-difference $\Theta$ depending on $N_{JJ}$ for the four JTWPAs. 
		(d) Gain responses of the \emph{Sando} JTWPAs with $N_{JJ}=1998$ calculated using the analytical approximation (dashed lines with circles) versus numerically integrating the system's differential equations (solid lines). 
		(e) Normalized wavenumbers of the signal (solid lines), pump (dotted lines) and idler (dashed lines).
		(f) Total phase-difference for the four sub-unit-cell sizes studied.
		The vertical dashed red line in (a), (b) and (c) indicates the devices represented in (d), (e) and (f).}
	\label{fig:pump_dep}
\end{figure*}
\subsection{Gain dependence on the number of junctions} 

Nonlinear amplification processes rely on the interchange of energy between different modes, where a weaker signal at a given frequency is enhanced by removing energy from a stronger mode, the pump. 
Naturally, one must be wary of reaching the point where the desired signal is drained by other modes and pump-depletion occurs~\cite{Yaakobi2013, Yaakobi2013Err}.
To avoid this, we operate the device in the strong-pump regime, where the pump mode is orders of magnitude larger than the other modes and does not suffer substantial depletion, and design a device with a number of JJs such that the signal is in the amplification regime.
To analyze this, in Fig.~\ref{fig:pump_dep} we numerically solve Eqs.~\eqref{eq:sando_DE_final} up to 200,000 JJs for \emph{Sando} JTWPAs with four different sub-unit-cell sizes: $x_0=3/4$ (cyan), $x_0=1/2$ (blue), $x_0=1/3$ (gray) and $x_0=1/5$ (orange).
Figure~\ref{fig:pump_dep}(a) shows the signal (solid lines), pump (dotted lines) and idler (yellow dashed lines) modes at $f=f_p$ depending on $N_{JJ}$ in logarithmic scale.
The signal and idler show amplification and deamplification regimes, which occur periodically with the $N_{JJ}$. This periodicity is clearly marked by the change of the total phase $\Theta$ [Eq.~\eqref{eq:sando_DE_final_th}] shown in Fig.~\ref{fig:pump_dep}(c).
Even though there is energy exchange between modes, the deamplification of the pump mode is negligible as it can be seen by the dotted red line in Fig.~\ref{fig:pump_dep}(a). If the pump and signal mode would be of the same order of magnitude, the regimes of pump deamplification and amplification would be non-negligible.
To better visualize signal amplification, in Fig.~\ref{fig:pump_dep}(b) we depict signal gain in decibels for the four sub-unit-cell sizes. Here it is clear that the benefit of tailoring the sub-unit-cell size is that maximum possible gain can be achieved using less junctions. For this ideal JTWPA, the maximum gain is 40 dB and would require 40,000 JJs, this is around ten times more JJs than currently fabricated JTWPAs. 
To analyze a more realistic setup, we limit ourselves to a JTWPA with 1998 JJs, devices with this number of JJs are currently routinely fabricated~\cite{White2015, Macklin2015, Hidaka2021, Sandbo2025arx}


Figures~\ref{fig:pump_dep}(d), (e) and (f) show the frequency dependence of these four \emph{Sando} JTWPAs in the strong-pump regime, evaluated at $N_{JJ}=1998$ and marked in Figs.~\ref{fig:pump_dep}(a), (b) and (c) with vertical red dashed lines.
Figure~\ref{fig:pump_dep}(d) shows the signal gain versus frequency for four different sub-unit-cell sizes, solid lines represent the solution obtained by numerical integration of Eqs.~\eqref{eq:sando_DE_final}, while the black dashed lines with circles show the strong-pump analytical solution [Eq.\eqref{eq:Sando_Gs}].
We can see a perfect agreement between analytical and numerical solutions for all four designs, validating the strong-pump approximation for ${N_{JJ}=1998}$.
Figure~\ref{fig:pump_dep}(e) shows the normalized wavenumber of the signal (solid lines), idler (dashed lines) and pump (dotted lines) modes for the different sub-unit-cell sizes.
Finally, Fig.~\ref{fig:pump_dep}(f) shows the total phase-difference depending on the operating frequency, interestingly the maximum gain is not achieved at the sub-unit-cell size which minimizes $\Delta k$, we will discuss this later on.

\subsection{Pump power and frequency}

\begin{figure}
	\centering
	\includegraphics{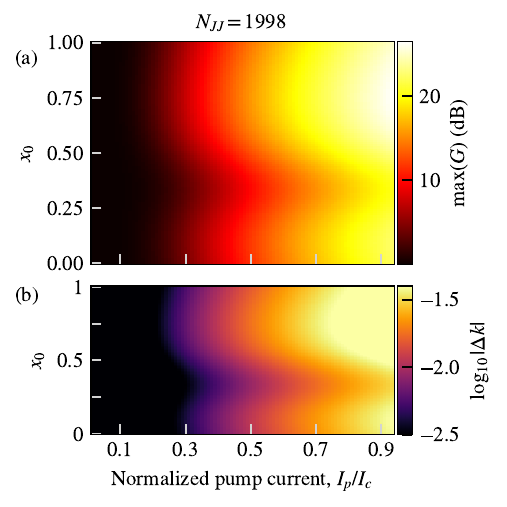}
	\caption{Gain (a) and phase-difference (b) depending on the normalized pump current $I_p/I_c$ and sub-unit-cell size $x_0$ with $N_{JJ}=1998$, $I_s/I_p=10^{-5}$ and $I_i/I_p=10^{-8}$.}
	\label{fig:pump_power}
\end{figure}

Figure~\ref{fig:pump_power} shows the maximum gain and the phase-difference dependence on the pump-power and sub-unit-cell size. 
We see the usual gain increase with pump-power, similarly the nonlinear interaction  dependence on $x_0$ is also enhanced for high pump-powers.

\begin{figure}
	\centering
	\includegraphics{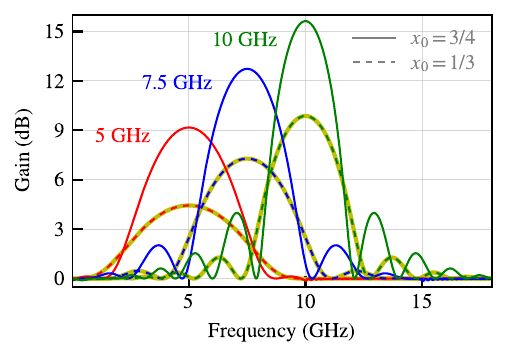}
	\caption{Gain spectrum of a \emph{Sando} JTWPA for two sub-unit-cell sizes $x_0=3/4$ (solid lines) and $x_0=1/3$ (dashed lines) operated at different pump frequencies: $f_p=5\text{ GHz}$ (red), $f_p=7.5\text{ GHz}$ (blue) and ${f_p=10\text{ GHz}}$ (green).
		Yellow solid lines depict a JTWPA defined by a typical single junction unit cell with $N_{JJ}=1998$. 
	}
	\label{fig:gain_pumps}
\end{figure}

We have seen that by changing the relative distance between the nonlinear elements we can achieve an enhancement of the device performance independent of engineering phase-matching.
Similarly to the \emph{Kerr-reversal} method~\cite{Ranadive2022}, the \emph{Sando} unit cell offers gain enhancement independent of the pump frequency. While bandwidth remains approximately the same.
Figure~\ref{fig:gain_pumps} depicts the gain spectrum of the same JTWPA pumped at different frequencies:
$f_p=5\text{ GHz}$ (red lines), $f_p=7.5\text{ GHz}$ (blue lines) and ${f_p=10\text{ GHz}}$ (green lines). For two different sub-unit-cell size; $x_0=3/4$ (solid lines) and $x_0=1/3$ (dashed lines).
The sub-unit-cell size $x_0=3/4$ outperforms $x_0=1/3$ for all pump frequencies, with a stronger improvement at lower pump frequencies, for instance, at ${f_p=5 \text{ GHz}}$ the ratio between maximum gains,  ${\max(G_{3/4})/\max(G_{1/3}) = 2.07}$, while at ${f_p=10 \text{ GHz}}$ it decreases slightly to  ${\max(G_{3/4})/\max(G_{1/3}) = 1.58}$.

\subsection{Gain saturation and 1-dB compression point}

Understanding the power range at which we can operate a nonlinear amplifier is critical for design optimization.
The $\text{1-dB compression point}$ elucidates up to which input signal power the device presents a constant gain and basically determines the power range where we can operate the device reliably. 

\begin{figure}
    \includegraphics{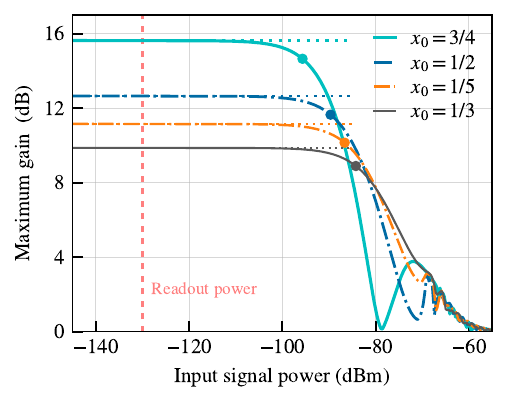}
    \caption{Maximum gain depending on the input signal power for JTWPAs with different sub-unit-cell sizes: $x_0=3/4$ (cyan solid line), $x_0=1/2$ (blue dash-dotted line), $x_0=1/3$ (gray solid line) and $x_0=1/5$ (orange dash-dotted line). The pump power is fixed at $-65.05\text{ dBm}$ ($I_p/I_c=0.5$), $I_i/I_p=10^{-8}$ and ${N_{JJ}=1998}$. The circles indicate the $\text{1-dB compression point}$, i.e. the point at which the maximum gain decreases by 1 dB. The horizontal dotted lines are obtained using the strong-pump approximation.} 
    \label{fig:gain_saturation}
\end{figure}

Figure~\ref{fig:gain_saturation} shows the maximum signal gain versus the input signal power in dBm of \emph{Sando} JTWPAs with different sub-unit-cell sizes. 
In the previous section, we showed that by changing the relative distance between the nonlinear elements, we can get an increase in gain. However, this comes with the cost of reducing its dynamic range.
For instance, due to the stronger nonlinearity of the sub-unit cell with $x_0=3/4$, its dynamic range is the narrowest showing a $\text{1-dB compression point}$ at around $-96.6 \text{ dBm}$, while for equal sub-unit-cell sizes, $x_0=1/3$, it occurs at around $-84.6\text{ dBm}$.
This narrower dynamic range has to be taken into account because it can reduce the device usability. 
For instance, if we are reading out a superconducting qubit using a dispersively coupled resonator at 10 GHz with linewidth ${\kappa = \left(30 \text{ ns} \right)^{-1}}$~\cite{Sank2025} and assuming a single photon in the resonator, then the readout power is about $-130 \text{ dBm}$ (marked by a vertical red dotted line in Fig.~\ref{fig:gain_saturation}), thus one would like to increase its $\text{1-dB compression point}$ to higher powers for multiple qubit readout.
Also, it has been shown that operating the JTWPA near the $\text{1-dB compression point}$ increases intermodulation products, which are detrimental for the amplifier's performance~\cite{Remm2023}.

\subsection{Sub-unit-cell size effects}

\begin{figure*}
	\includegraphics{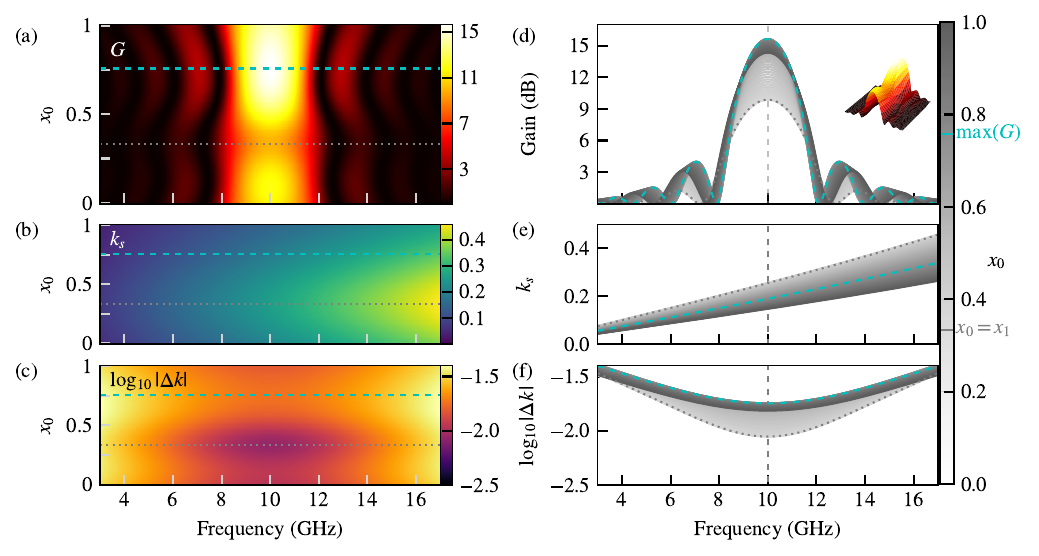}
	\caption{JTWPA's performance depending on the signal frequency and the sub-unit-cell size $x_0$. (a) Gain in dB pumped at $f=10 \text{ GHz}$ with normalized pump current $I_p/I_c=0.5$. (b) Signal mode wavenumber and (c) phase difference.
	For easier visualization, (d)-(f) show cross-sections of (a)-(c).
	Cyan dashed lines mark the sub-unit-cell size that maximizes gain ($x_0 = 3/4$), while the gray dotted lines denote the minimum gain ($x_0=1/3$).
	}
	\label{fig:sando_gain}
\end{figure*}

We showed the validity of the strong-pump approximation to simulate the signal gain with $N_{JJ}=1998$ in Fig.~\ref{fig:pump_dep}(d).
Using Eq.~\eqref{eq:Sando_Gs}, 
we investigate the signal gain and bandwidth dependence on the sub-unit-cell size $x_0$.
Figures~\ref{fig:sando_gain}(a), (b) and (c) show heatmaps of gain, signal-mode wavenumber and phase-difference depending on $x_0$ and signal frequency. To better visualize these dependencies, Figs.~\ref{fig:sando_gain}(d), (e) and (f) show cross-sections of (a), (b) and (c), respectively.
While the effects on the wavenumber and phase-difference, [Fig.~\ref{fig:sando_gain}(b) and (c), respectively], are minor, the JJs distance's effect on gain and bandwidth is substantial due to the dependence of the nonlinear scaling factor [Eq.~\eqref{eq:sando_NL_factor}] on $x_0$.
Figure~\ref{fig:sando_gain}(a) and (d) show two local maxima as a function of $x_0$ and a minimum corresponding to $x_0=1/3$ (equal sized sub-unit cells). The maximum gain at $x_0=2/3$ presents over $ 20 \text{ dB}$ improvement compared to $x_0=1/3$ with a bandwidth wider than $ 5 \text{ GHz}$ above the $20 \text{ dB}$ mark.
These results only show the capabilities of this approach and could be improved by further optimizing the device and input parameters, e.g. increase the number of JJs per unit cell (which interact to each other), minimizing the phase-difference by using an LC-resonator~\cite{OBrien2014}, or other engineering techniques~\cite{Peng2022_PRX, Gaydamachenko2022}.

\begin{figure}
	\includegraphics{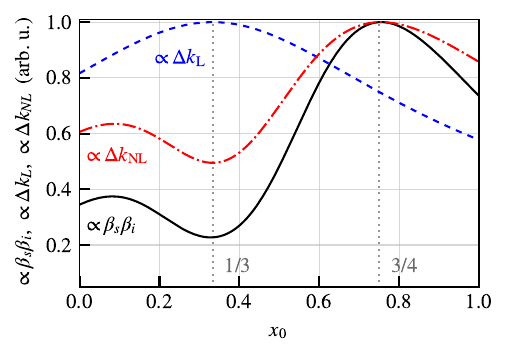}
	\caption{The sub-uni-cell size $x_0$ dependent terms of  $\Delta k_{L}$ (blue dashed line), $\Delta k_{NL}$ (red dash-dotted line) and $\beta_s\beta_i$ (black solid line). For easy comparison these terms have been normalized.}
	\label{fig:proportional}
\end{figure}

To understand the role of the scaling factors in the JTWPA gain and why maximum gain occurs at a $\Delta k$ saddle point in the $(x_0, \omega_s)\text{-space}$, we analyze the strong-pump gain equation, Eq.~\eqref{eq:sando_g}. 
The gain scales with $\beta_s\beta_i - (\Delta k /2)^2$, thus the usual strategy to optimize gain is to achieve phase-matching ($\Delta k=0$) by design engineering~\cite{OBrien2014, Planat2020, Ranadive2022}.
However, in the \emph{Sando} model we have an extra degree of freedom, $x_0$, and both terms $\beta_s\beta_i$ and $(\Delta k /2)^2$ have different dependencies on $x_0$. 
Fixing the operating frequencies and pump strength, we find that $\beta_s\beta_i \propto  \heartsuit^{-5} \left(\diamondsuit^{\text{NL}}_{s;p,p,-i}\right)^2 $, and the phase-difference linear and nonlinear terms scale as
$\Delta k_{\text{L}} \propto \heartsuit^{-1/2}$  and
$	\Delta k_{\text{NL}} \propto \heartsuit^{-5/2} \diamondsuit^{\text{NL}}_{p;p,p,-p}$ respectively, where $\Delta k = \Delta k_{\text{L}} + \Delta k_{\text{NL}}$.
Therefore, it is not possible to maximize $\beta_s\beta_i$ while minimizing $\Delta k$ as a function of $x_0$, since $\beta_s\beta_i$ and $\Delta k_{\text{NL}}$ scale proportionally. In fact, if $\Delta k_{\text{L}}=0$ then 
$\diamondsuit^{\text{NL}}_{s;p,p,-i}=\diamondsuit^{\text{NL}}_{p;p,p,-p}$, so the gain follows $g \propto \heartsuit^{-5/2} \diamondsuit^{\text{NL}}_{p;p,p,-p}$.
Figure~\ref{fig:proportional} shows the $x_0$ dependent terms of  $\Delta k_{\text{L}}$ (blue dashed line), $\Delta k_{\text{NL}}$ (red dash-dotted line) and $\beta_s\beta_i$ (black solid line). As expected, $\beta_s\beta_i$ and $\Delta k_{\text{NL}}$ have the same dependency on $x_0$, also agreeing with the trend seen for the JTWPA gain in Fig.~\ref{fig:sando_gain}.

Physically the propagation of each phase mode depends on $k(\omega)$, thus at a given distance the phase of each mode evolves differently. In a nonlinear medium the energy exchange between modes depends on their phases Eq.~\eqref{eq:sando_DE_final}, thus by changing the relative distance between junctions we can study which spacing creates the most favorable phase off-set when the wave reaches the nonlinear element.

\subsection{Resonant phase matching}
\begin{figure}
	\centering
	\includegraphics[width=\linewidth]{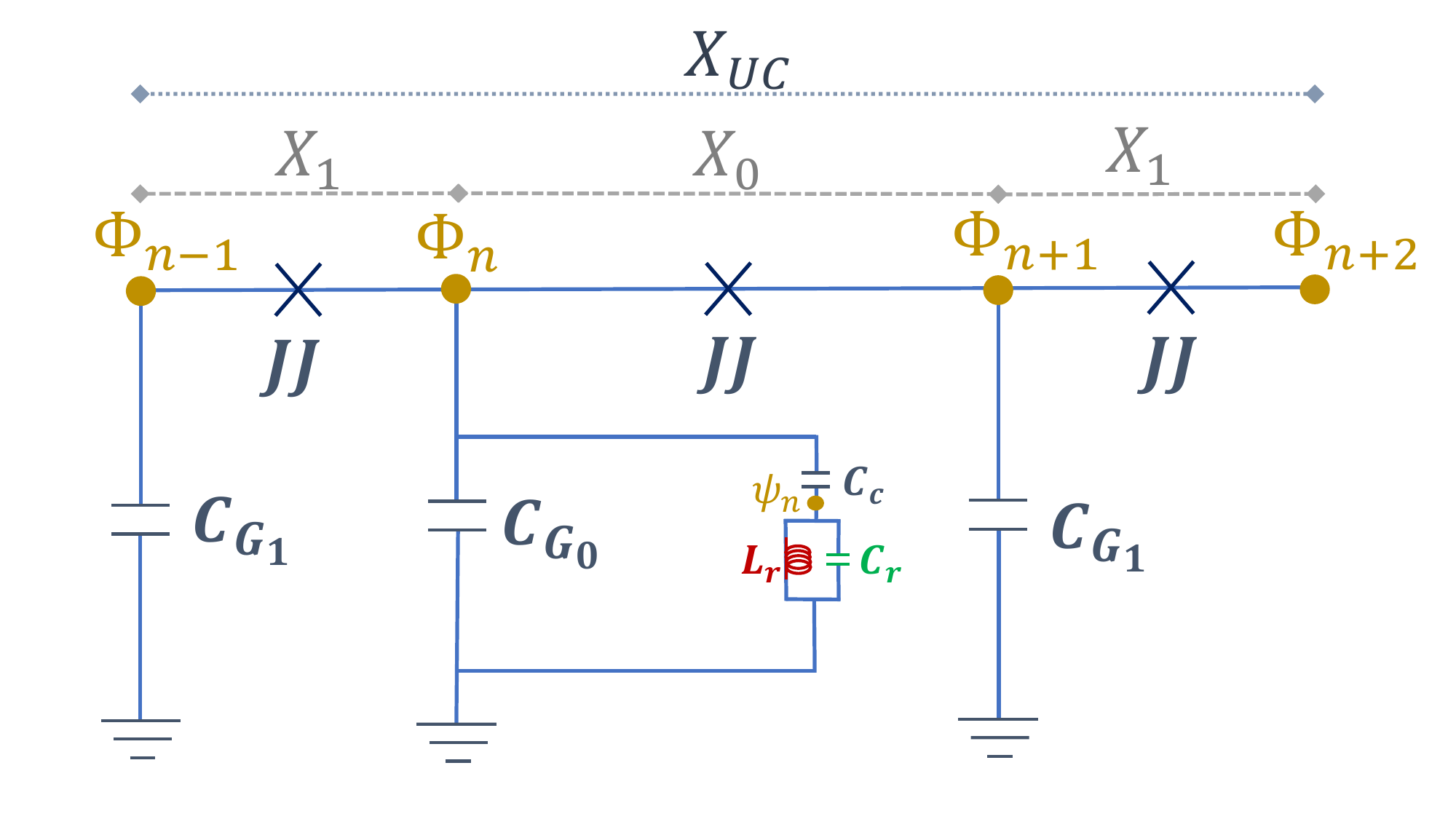}
	\caption{Sando unit cell with a resonator in the central sub-unit cell.
		The total unit cell length is $X_{UC}$ and each sub-unit cell has a length defined by $X_{0,1}$, the crosses represent Josephson junctions (defined by $I_c$ and $C_J$) and each sub-unit cell contains a ground capacitance $C_{G_{0,1}}$. The gold circles indicate the node fluxes $\Phi_n, \, \psi_n$ between branches. The central sub-unit cell has an LC-resonator characterized by an inductance $L_r$, capacitance $C_r$ and coupling capacitance $C_c$. 
	}
	\label{fig:UC-diagram_res}
\end{figure}

\begin{figure*}
	\includegraphics{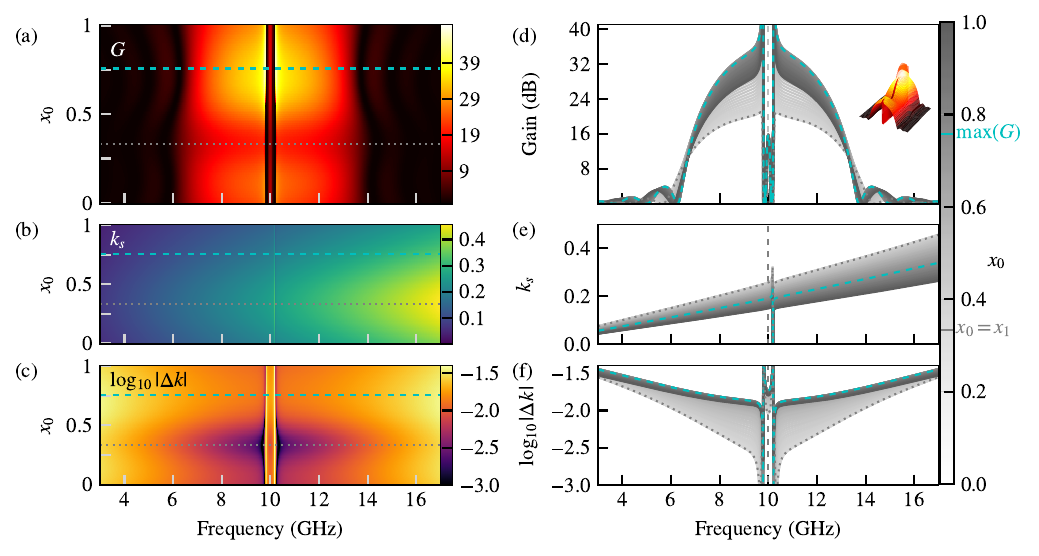}
	\caption{RPM-JTWPA's performance dependence on signal frequency and $x_0$ for a \emph{Sando} unit cell with resonant frequency $f_r=10.2$ GHz ($C_r=6$ pF, $L_r=40.6$ pH, $C_c=20$ fF, $C_{G_0}=6.33$ fF and $C_{G_1}=26.3 \text{ fF}$). (a) Signal gain in dB pumped at $f=10\;$GHz with normalized current $I_p/I_c=0.5$. (b) Signal mode wavenumber and (c) phase difference at different sub-unit-cell sizes.
		For easier visualization, (d)-(f) show cross-sections of (a)-(c), respectively.
		Cyan dashed lines mark the sub-unit-cell size that maximizes gain ($x_0 = 3/4$), while the gray dotted lines denote the minimum gain ($x_0=1/3$).}
	\label{fig:sando_gain_LCres}
\end{figure*}
The method introduced in this work can be easily extended to unit-cells with LC-resonators in order to get resonant phase matching~\cite{OBrien2014}.
For instance, if we include a resonator in the sub-unit cell in the middle (Fig.~\ref{fig:UC-diagram_res}), after minimizing the action we would obtain similar equations than Eqs.~\eqref{eq:sando_DE_final}, however in this case the linear \emph{Sando} scaling factor $\clubsuit$ would include the resonator effective capacitance,
\begin{align}
  \clubsuit (\omega) := \left[ c_{G_0} + c_{LC}(\omega) \right] + 2c_{G_1}\, ,
\end{align}
where 
\begin{align}
	c_{LC}(\omega) := c_c \frac{l_rc_r\omega^2 - 1}{l_r\left(c_c + c_r \right) \omega^2 -1} \, ,
\end{align}
with $c_c$, $c_r$ and $l_r$ the normalized coupling capacitance and resonator's capacitance and inductance, respectively.

Note that the resonator effective capacitance depends on the operating frequency $\omega$. As expected,  for unit cells with resonators, the scaling factor $\clubsuit$ also becomes frequency dependent.

Figure~\ref{fig:sando_gain_LCres} shows the effect of a \emph{Sando} unit cell in a JTWPA with an LC-resonator designed to maximize resonant phase matching (RPM). The LC-resonator is located in the sub-unit cell in the middle and has a resonant frequency of $10.2$ GHz defined by $C_r=6$ pF and $L_r=40.6$ pH, the coupling capacitance is $C_c=20$ fF and the zeroth order capacitance $C_{G_0}=6.33$ fF, so the transmission line impedance remains $Z_0=50 \, \Omega$.
The benefits of RPM-JTWPA are still present when a \emph{Sando} unit cell is used, thus combining the strengths of both techniques can further extend the device bandwidth and increase its gain.
For instance, in the 1 - 20 GHz regime the propagating waves have wavelength of a few millimeters, thus a unit cell size $X_{UC} < 100 \, \upmu $m should satisfy the continuous and SVE approximations, Eqs.~\eqref{eq:cont_appr} and \eqref{eq:phi_an} respectively. Current fabrication technologies can routinely fabricate $10 \, \upmu$m unit cells~\cite{Macklin2015, Hidaka2021}, assuming this to be the smaller sub-unit-cell size $X_1$, then for the optimal configuration ($X_0/X_{UC}=0.75$) $X_0=60 \, \upmu$m, and the total unit-cell size $X_{UC}=80 \, \upmu$m, which should satisfy this model theoretical requirements and, following the ideal results presented in Fig.~\ref{fig:sando_gain_LCres}(a) and (d), give a 72\% gain increase at 9 GHz compared to a typical unit cell (equal spacing).
A more conservative design would be using a sub-optimal ratio $X_0 / X_{UC}=0.5$, in this case $X_0=20 \, \upmu$m, $X_1=10 \, \upmu$m and $X_{UC}=40 \, \upmu$m, and it would give a 36\% gain increase at 9 GHz.

Note that as shown in Fig.~\ref{fig:pump_dep}(a) and (b), the maximum gain could also be achieved with a typical unit-cell size, but using about double number of Josephson junctions. The \emph{Sando} unit cell requires less JJs, but the total unit-cell size becomes larger, therefore if significant loss per unit length is present the benefits of this approach might be capped.

\section{Conclusion}\label{sec:Con}

In this work we have introduced a novel theoretical model which allows to investigate
how sub-wavelength spacing between Josephson junctions affects wave mixing. By including three Josephson junctions within a unit cell, we can analyze the effective response due to their nonlinear interaction.
Here, we focus on a particular design, the \emph{Sando} unit cell, in which the size of two sub-unit cells are equal, while the other one is allowed to have arbitrary relative size (Fig.~\ref{fig:unit_cell}).

In Fig.~\ref{fig:sando_gain}, we have shown that by changing the relative distance between junctions we can obtain a larger gain and wider bandwidth. Moreover, it exists an optimal ratio which maximizes both. 
However, this increase in gain comes at the cost of narrower dynamic range, Fig.~\ref{fig:gain_saturation}.
On the other hand, due to the stronger effective nonlinearity of the \emph{Sando} unit cell, one can achieve devices with larger gain and bandwidth with less Josephson junctions, Fig.~\ref{fig:pump_dep}, which can be beneficial for the device fabrication and footprint.
By including a resonator to achieve resonant phase matching, we can further increases gain and bandwidth, but with the usual drawbacks of having a fixed operating pump frequency and a stopband in the gain spectrum. Our model predicts an ideal gain increase of 70\% for the optimal sub-unit-cell size, producing gain over 29 dB on a 4 GHz bandwidth.

To implement this method to experimental designs, one must be wary of the sub-wavelength distance requirement between Josephson junction. This can become a handicap for fabricating a device at the optimal sub-unit-cell size, however even designs at sub-optimal ratios should present substantial performance improvement Fig.~\ref{fig:sando_gain_LCres}.

Even though we have focused on the \emph{Sando} unit cell, the mathematical framework introduced here can be extended to arbitrary sized unit cells, also allowing for any number of Josephson junctions within the unit cell. In future work, we plan to investigate these cases, also including the effects of circuit parameter spreads~\cite{Kissling2023} and stochastic variations in the sub-unit cell sizes along the JTWPA.

To summarize, the theory presented in this work points to complimentary engineering methods to maximize JTWPA performance, and opens the field of research into sub-unit-cell engineering in order to optimize the effective nonlinear strength of these quantum amplifiers.

%

\section*{Acknowledgments}
This paper was based on results obtained from a
project, JPNP16007, commissioned by the New Energy
and Industrial Technology Development Organization
(NEDO), Japan.

\section*{Author Contribution Statements}
{\bf M.A.G.L.}: Conceptualization, Formal Analysis, Methodology, Investigation, Software and Writing - Original Draft.
{\bf T.Y.}: Investigation and Writing - Review \& Editing.
{\bf Y.N.}: Investigation and Writing - Review \& Editing.
{\bf Y.U.}: Investigation and Writing - Review \& Editing.
{\bf J.C.}: Formal Analysis and Writing - Review \& Editing.
{\bf K.I.}: Investigation, Writing - Review \& Editing, Project Administration and Funding Acquisition.

\appendix

\section{Mathematical model}

The explicit expression of $d\Theta /dx$ (appearing in Eq.~\eqref{eq:sando_DE_final_th}) is
\begin{widetext}
\begin{subequations}\label{eq:dthetadx}
\begin{align}
 \frac{\partial \Theta}{\partial x} =& \Delta k_L + 2\frac{\partial \theta_{p} }{\partial x} 
 - \frac{\partial \theta_s }{\partial x} - \frac{\partial \theta_i }{\partial x}  \\
=& \Delta k_L + \frac{\diamondsuit^{\text{NL}}_{\Delta k} \, k_{p}^2k_s k_i \xi_{p}^2\xi_s\xi_i}{2^4 \clubsuit }
\left[
4\frac{ \left(  k_s + k_i - k_{p} \right) }{\xi_{p}^2\omega_{p}^2  } 
- \frac{  \left(  2k_p - k_i \right) }{ \xi_s^2 \omega_s^2  } 
- \frac{ \left(  2k_p - k_s \right) }{ \xi_i^2 \omega_i^2 } \right] \cos \Theta \nonumber \\
& + \frac{k_{p}^3 \diamondsuit^{\text{NL}}_{\text{self}} }{ 2^3 \omega_{p}^2  \clubsuit} \Big( k_{p}^2  \xi_{p}^2  
+ 2 k_s^2  \xi_s^2  + 2 k_i^2  \xi_i^2 
\Big) 
   - \frac{k_s^3 \diamondsuit^{\text{NL}}_{\text{self}} }{ 2^4 \omega_s^2 \clubsuit  } \Big(
 k_s^2  \xi_s^2 
+ 2 k_{p}^2  \xi_{p}^2  
+ 2 k_i^2  \xi_i^2 
\Big) 
 - \frac{k_i^3 \diamondsuit^{\text{NL}}_{\text{self}} }{ 2^4 \omega_i^2  \clubsuit  } \Big(
k_i^2  \xi_i^2 
+ 2 k_{p}^2  \xi_{p}^2   
+ 2 k_s^2  \xi_s^2 
 \Big) 
\, ,
\end{align}
\end{subequations}
\end{widetext}
where $\diamondsuit^{\text{NL}}_{\text{self}}:= \diamondsuit^{\text{NL}}_{n;m_{1,2,3}}$ such that ${k_{m_1} + k_{m_2} + k_{m_3} - k_n = 0}$.

\subsection{Strong-pump approximation}
\label{app:strongpump}

In this section, we will make use of the strong-pump approximation to obtain an analytical expression for the signal and idler.

In the strong-pump approximation, the pump modes are assumed to be much larger than the idler and signal modes, i.e. $|A_{p_1, p_2}| \gg |A_{s,i}|$. 
For the particular case of degenerate pumps, we can express the propagating wave as,
\begin{align}
	\phi(x, \tau) = & \frac{1}{2} \big[ A_{p}(x) e^{\im \psi_p}  
	+ A_{s}(x) e^{\im \psi_s} + A_{i}(x) e^{\im \psi_i} + c.c. \big] \, ,\label{eq:sando_strong_pump_wave}
\end{align}
where $\psi_m := k_mx - \omega_m \tau$.
Then, expanding Eq.~\eqref{eq:sando_DE1} and using the strong-pump approximation, i.e. neglecting contributions from $A_{s,i}$, we obtain
\begin{align}\label{eq:DEpump_sp}
	\frac{\partial A_{p}}{\partial x} = \im \frac{k_{p}^5\diamondsuit^{\text{NL}}_{\text{self}}}{2^4\omega_{p}^2 \clubsuit}  A_{p}  |A_{p}|^2  \, .
\end{align}

Expressing $A_p(x)=a_{p}(x) e^{\im \alpha_{p} x}$ in Eq.~\eqref{eq:DEpump_sp} and assuming degenerate pumps with equal initial conditions $a_{p}(0)=a_p$ we obtain,
\begin{align}
	\frac{\partial a_{p}(x)}{\partial x} = 0 \, , 
\end{align}
which defines a rotating wave with constant amplitude $a_p$ and phase 
\begin{align}\label{eq:sando_gain_factors_a_p}
	\alpha_{p} := \frac{ k_{p}^5 \diamondsuit^{\text{NL}}_{\text{self}} }{2^4\omega_{p}^2\clubsuit}  a_{p}^2   \, .
\end{align}

Now we can proceed to analyze the signal and idler modes, applying the strong-pump approximation to Eq.~\eqref{eq:sando_DE1} we obtain
\begin{subequations}
	\label{eq:strong-pump_Asi}
\begin{align}
	\frac{\partial A_s}{\partial x} =& \im \frac{k_s }{2^4\omega_s^2 \clubsuit } \Big( 
	\diamondsuit^{\text{NL}}_{\Delta k} k_{p}^2 k_i \left( 2k_{p} - k_i \right) A_{p}^2 A_i^* e^{\im \Delta k_L  x}
	\nonumber \\
	& + 2 k_s^2A_s \diamondsuit^{\text{NL}}_{\text{self}}  k_{p}^2 |A_{p}|^2 \Big) \, ,
	 \\
	\frac{\partial A_i}{\partial x} =& \im \frac{k_i }{2^4\omega_i^2 \clubsuit } \Big( 
	\diamondsuit^{\text{NL}}_{\Delta k} k_{p}^2 k_s \left( 2k_{p} -  k_s\right) A_{p}^2 A_s^* e^{\im \Delta k_L  x}
	\nonumber \\
	& + 2 k_i^2 A_i \diamondsuit^{\text{NL}}_{\text{self}}  k_{p}^2 |A_{p}|^2 \Big) \, , 
\end{align}
\end{subequations}
These equations can be simplified by expressing the complex amplitudes as
\begin{align}
	A_n(x) &= a_n(x) e^{\im \alpha_n x} \, ,
\end{align}
where
\begin{subequations}\label{eq:sando_gain_factors_a_si}
\begin{align}
	\alpha_s := \frac{k_s^3 k_p^2  \diamondsuit^{\text{NL}}_{\text{self}}}{2^3 \omega_s^2 \clubsuit }  a_{p}^2  \, , \quad
	\alpha_i  := \frac{k_i^3 k_p^2 \diamondsuit^{\text{NL}}_{\text{self}}}{2^3 \omega_i^2 \clubsuit }   a_{p}^2  \, ,
\end{align}
\end{subequations}

Then, the Eqs.~\eqref{eq:strong-pump_Asi} can be rewritten as,
\begin{subequations}\label{eq:reduced_DE}
	\begin{align}
		\frac{\partial a_s}{\partial x} &= \im \beta_s a_i^* e^{\im \Delta k x}  \, , \\
		\frac{\partial a_i}{\partial x} &= \im \beta_i a_s^* e^{\im \Delta k x} \, ,
	\end{align}
\end{subequations}
where we have defined,
\begin{align}\label{eq:sando_Dk}
	\Delta k & := 2k_p - k_i - k_s + 2\alpha_p -\alpha_s -\alpha_i \, ,
\end{align}	
\begin{subequations}\label{eq:sando_gain_factors_b}
\begin{align}
	\beta_s & :=  \frac{k_p^2 k_i k_s  \left( 2k_{p} - k_i \right) \diamondsuit^{\text{NL}}_{\Delta k}}{2^4\omega_s^2  \clubsuit }    a_p^2 \, , \\
	\beta_i &:=  \frac{k_p^2 k_i k_s  \left( 2k_{p} - k_s \right) \diamondsuit^{\text{NL}}_{\Delta k}}{2^4\omega_i^2  \clubsuit }  a_p^2 \, .
\end{align}
\end{subequations}

By solving Eqs.~\eqref{eq:reduced_DE}, we obtain the signal mode
\begin{align}
	a_s(x) =& \bigg\{ \left[ \cosh (gx) - \im \frac{\Delta k}{2g} \sinh(gx) \right] a_{s}(0) \nonumber \\
	& - \im \frac{\beta_s}{g}\sinh(gx) a_{i}^*(0) \bigg\} e^{\im \frac{\Delta k}{2}x} \, .
\end{align}

Then, the magnitude of the signal mode at an arbitrary $x$, and assuming that $|a_{i}(0)|\approx 0$,
\begin{equation}\label{eq:signal_mode}
	|a_s(x)|^2 = \left[ \cosh^2(gx) + \left(\frac{\Delta k}{2g} \right)^2 \sinh^2(gx) \right] |a_{s}(0)|^2 \, ,
\end{equation}
where the gain coefficient is
\begin{align}\label{eq:sando_g}
g &= \sqrt{\beta_s \beta_i - \left( \frac{\Delta k}{2} \right)^2 } \, .
\end{align}

From Eq.~\eqref{eq:signal_mode}, the gain (Eq.~\eqref{eq:Sando_Gs}) is immediately obtained.

Note that when numerically calculating the gain using Eqs.~\eqref{eq:signal_mode}-\eqref{eq:sando_g} one must be careful at zero detuning, since $g=0$ and Eq.~\eqref{eq:signal_mode} can run into a numerical issue. To avoid any mishaps, at zero detuning we define Eq.~\eqref{eq:signal_mode} using limits, 
\begin{align}\label{eq:signal_mode_limit}
	\lim_{g\rightarrow 0}\frac{|a_s(x)|^2}{|a_{s}(0)|^2} &= 	\lim_{g \rightarrow 0}\left[ \cosh^2(gx) + \left(\frac{\Delta k}{2g} \right)^2 \sinh^2(gx) \right]  \nonumber \\
	&=  1 + \left(\frac{\Delta k \, x}{2 } \right)^2 
	\, ,
\end{align}
where we have used that $ \lim_{g \rightarrow 0} \sinh^2(g x) = (g x)^2$.

\subsection{LC-Resonator}
If we add a resonator in the central sub-unit-cell as represented in Fig.~\ref{fig:UC-diagram_res} an extra term will appear in the unit cell Lagrangian density (Eq.~\eqref{eq:sando_Lagrangian2}),
\begin{align}
	\mathcal{L} = \mathcal{L}_1 + \mathcal{L}_2 + \mathcal{L}_3 + \mathcal{L}_4 \, ,
\end{align}
where the last term in the Lagrangian density is
\begin{align}
	\mathcal{L}_4 =& \frac{E_J}{2} \Big\{  c_{c} \left[ \dot{ \psi}(x, \tau) - \dot{ \phi}(x, \tau) \right]^2 \nonumber \\&
	+ c_{r}  \dot{ \psi}(x, \tau)^2 
	- \frac{1}{l_r}  \psi(x, \tau)^2 \Big\}\, ,
\end{align}
where $\psi(x, \tau)$ is an auxiliary node flux at the resonator, see Fig.~\ref{fig:UC-diagram_res}.

Then, the variation of $\mathcal{L}_4$ is
\begin{align}
	\delta \mathcal{L}_4 =& E_J \Big\{  c_{c} \left[ \dot{ \psi}(x, \tau) - \dot{ \phi}(x, \tau) \right] 
	\left[ \delta \dot{ \psi}(x, \tau) - \delta \dot{ \phi}(x, \tau) \right] \nonumber \\&
	+ c_{r}  \dot{ \psi}(x, \tau) \, \delta  \dot{ \psi}(x, \tau)
	- \frac{1}{l_r}  \psi(x, \tau) \, \delta \psi(x, \tau)   \Big\} \, ,
\end{align}
from which its action can be found
\begin{widetext}
	\begin{subequations}
		\begin{align}
			\frac{\delta \mathcal{S}_4 }{ E_J}=& \frac{1}{ 2\pi } \iint dx d\tau \Big\{ c_c \iint d\omega d\omega' (-\omega \omega') e^{-\im (\omega + \omega')\tau}
			\left[  \hat{\psi}(x, \omega) -  \hat{\phi} (x, \omega) \right] 
			\Big[ \delta \hat{\psi} (x, \omega')  - \delta   \hat{\phi} (x, \omega')  \Big] \Big\} \nonumber \\&
			+ \frac{1}{ 2\pi }\iint dx d\tau \Big\{ c_r \iint d\omega d\omega' (-\omega \omega') e^{-\im (\omega + \omega')\tau} \hat{\psi} (x, \omega) \,  \delta \hat{\psi}(x, \omega')  \Big\} 
			- \iint dx d\tau  \frac{1}{l_r}  \psi (x, \tau) \, \delta  \psi(x,  \tau)  \label{eq:S4_a} \\
			=&  \frac{1}{\sqrt{2\pi}} \iiint dx d\tau d\omega \Big\{ c_c  \omega^2 
			\left[  \hat{\psi} (x, \omega) -  \hat{\phi} (x, \omega) \right] e^{-\im \omega \tau}
			\Big[ \delta  \psi (x, \tau)  - \delta \phi (x, \tau) \Big] \Big\} \nonumber \\&
			+ \frac{1}{\sqrt{2\pi}} \iiint dx d\tau  d\omega \, c_r \omega^2  \hat{\psi} (x, \omega) e^{-\im \omega \tau} \, \delta \psi(x, \tau)   
			- \frac{1}{\sqrt{2\pi}} \iint dx d\tau d\omega \frac{1}{l_r}  \hat{\psi} (x, \omega) e^{-\im \omega \tau} \, \delta \psi(x,  \tau)  \label{eq:S4_b}\\
			=& \frac{1}{\sqrt{2\pi}} \iiint dx d\tau d\omega \, c_c  \omega^2 \left[ \hat{ \phi} (x, \omega) - \hat{\psi} (x, \omega)\right] e^{-\im \omega \tau} \, 
		 \delta   \phi (x, \tau)   \nonumber \\&
		+ \frac{1}{\sqrt{2\pi}} \iiint dx d\tau  d\omega \left[ \left( c_c  \omega^2 + c_r \omega^2  -  \frac{1}{l_r}  \right) \hat{\psi} (x, \omega) - c_c  \omega^2  \hat{\phi} (x, \omega) \right] e^{-\im \omega \tau} \, \delta  \psi(x,  \tau)  \, , \label{eq:S4_c}	
		\end{align}
	\end{subequations}
\end{widetext}
where from Eq.~\eqref{eq:S4_a} to Eq.~\eqref{eq:S4_b} we have used the inverse-Fourier transformation and the \emph{delta distribution} definition, and from Eq.~\eqref{eq:S4_b} to Eq.~\eqref{eq:S4_c} we have factored together the terms depending on $\delta  \phi (x, \tau)  $ and $\delta  \psi (x, \tau)$.
In this case, the action of the system depends on two fields $\phi$ and $\psi$. Thus, the minimum trajectory of the total action, i.e. $\delta  S =0$, determines two equations: the part depending on the variation of $\phi$, $\delta  S (\phi) =0$, and the term depending on the variation of $\psi$, $\delta S (\psi)=0$.
By solving these two equations we will find the relationship between the two fields, expressed as
\begin{align}
	\hat{\psi} (x, \omega) = \frac{l_r c_c \omega^2}{l_r \left( c_r + c_c \right) \omega^2 -1 } \hat{\phi} (x, \omega) \, ,
\end{align}
and the scaling factor $\clubsuit$ will become
\begin{align}
	\clubsuit (\omega) := 2 c_{G_1} + \left[ c_{G_0} + c_{LC}(\omega) \right] \, ,
\end{align}
where due to the resonator, $\clubsuit (\omega)$ is now dependent on the operating frequency, with the resonator capacitance defined as
\begin{align}
	c_{LC}(\omega) := c_c \frac{l_r c_r \omega^2 -1}{l_r\left( c_r + c_c \right) \omega^2 -1} \, .
\end{align}

\bibliography{references.bib}

\end{document}